\documentclass[oneside]{amsart}
\usepackage{amsmath}
\usepackage{amssymb}
\usepackage[margin=3cm]{geometry}
\usepackage{hyperref}
\date{June 29, 2007}

\newtheorem{thm}{Theorem}[section]
\newtheorem{cor}[thm]{Corollary}
\newtheorem{prop}[thm]{Proposition}

\newtheorem{lemma}[thm]{Lemma}
\newtheorem{Remark}[thm]{Remark}
\newtheorem{theorem}{Theorem}[section]
\newtheorem{remark}[theorem]{Remark}

\renewenvironment{proof}{{\bf Proof:}}{\hfill$\square$\vskip.5cm}
\newenvironment{proofof}{}{\hfill$\square$\vskip.5cm}
\newcommand{\R}{\mathbb{R}}
\newcommand{\E}{\mathbb{E}}
\newcommand{\Z}{\mathbb{Z}}

\newcommand{\s}{\sigma}
\newcommand{\calF}{\mathcal{F}}
\newcommand{\sfH}{\mathsf{H}}
\newcommand{\sfi}{\mathsf{i}}
\newcommand{\sfj}{\mathsf{j}}
\newcommand{\sfn}{\mathsf{n}}
\newcommand{\sfJ}{\mathsf{J}}
\newcommand{\sfK}{\mathsf{K}}
\newcommand{\sfN}{\mathsf{N}}
\newcommand{\sfM}{\mathsf{M}}
\newcommand{\sfp}{\mathsf{p}}
\newcommand{\sfg}{\mathsf{g}}
\newcommand{\sfX}{\mathsf{X}}
\newcommand{\sfV}{\mathsf{V}}
\newcommand{\sfZ}{\mathsf{Z}}
\newcommand{\sfw}{\mathsf{w}}
\newcommand{\sfd}{\mathsf{d}}

\newcommand{\Var}{\operatorname{Var}}

\title{Some Observations for Mean-Field Spin Glass Models}

\author{Shannon Starr and Brigitta Vermesi}

\address{
Department of Mathematics\\
Hylan Building\\
University of Rochester\\
Rochester, NY 14627}

\date{\today}

\begin{document}

\markright{}

\begin{abstract}
We obtain bounds to show that the pressure of a two-body, mean-field spin glass
is a Lipschitz function of the underlying distribution of the random coupling constants,
with respect to a particular semi-norm.
This allows us to re-derive a result of Carmona and Hu, on the universality of the SK model,
by a different proof, and to generalize this result to the Viana-Bray model.
We also prove another bound, suitable when the coupling constants are not independent,
which is what is necessary if one wants to consider ``canonical'' instead of ``grand canonical''
versions of the SK and Viana-Bray models.
Finally, we review Viana-Bray type models, using the language of L\'evy processes,
which is natural in this context.
\end{abstract}

\maketitle

\section{Continuity of pressure with respect to the coupling distribution}

Let us consider a mean-field spin glass Hamiltonian of the form
$$
-\sfH_N(\s)\, =\, \sum_{i,j=1}^N \sfJ_N(i,j) \s_i \s_j + h \sum_{i=1}^N \s_i\, ,
$$
where the coefficients $\sfJ_N(i,j)$ are i.i.d.\ random variables, chosen from some distribution
represented by a cumulative distribution function, $F_N$, 
and $h \in \R$ is nonrandom.
The spins themselves are assumed to be Ising type spins, so that $\s = (\s_1,\dots,\s_N)$
lives in $\Omega_N = \{+1,-1\}^N$.
This general definition encompasses the 
SK model, where $F_N$ is the c.d.f., for a $\mathcal{N}(0,\beta^2/2N)$ random variable.
That is, all the $\sfJ_{N}(i,j)$'s are 
normal (Gaussian) random variables with mean $0$ and variance $1/2N$.
This definition also includes a nice version of a Viana-Bray type model,
wherein $F_N$ is the c.d.f.\ for a
Poissonized Gaussian random variable.
More specifically, for each $i$ and $j$, we have
$$
\sfJ(i,j)\, =\, \sum_{k=1}^{\sfK(i,j)} \sfg_{k}(i,j)\, ,
$$
where $\sfK(i,j)$ is a Poisson random variable, with mean $\alpha/N$,
and all the random variables $\sfg_{k}(i,j)$, for $k=1,2,\dots$,
are i.i.d.\ $\mathcal{N}(0,\beta^2)$ random variables.

Note that we write $\sfJ_N(i,j)$ and $\sfH_N(\s)$ using the {\sf sans serif} font to indicate that they are
random variables.
As usual, we also define the random partition function,
$$
\sfZ_N\, =\, \sum_{\s \in \Omega_N} e^{- \sfH_N(\s)}\, ,
$$
and the ``pressure'',
$$
\sfp_N\, =\, \frac{1}{N} \ln(\sfZ_N)\, .
$$
This random variable is not actually the pressure, which is only defined in the thermodynamic
limit.
But we will call it the ``random pressure'', anyway.
Let us write
$$
p_N(F_N)\, =\, \E[\sfp_N]\, ,
$$
which is the definition of the ``quenched pressure''.
Note that at this point we explicitly denote the underlying distribution $F_N$
for the coupling constants $(\sfJ_N(i,j)\, :\, 1\leq i,j\leq N)$.
This is because $p_N(F_N)$ is not random, but a function of $F_N$.

A very basic, but important bound is the following.
\begin{lemma}
\label{lem:basic}
Let $w_1,w_2,\dots,w_K \geq 0$ be given, and suppose $w_1+\dots+w_K \geq 0$.
Also, suppose $s_1,\dots,s_K \in [-1,1]$ are given.
Define a function $f : \R\to \R$ by
$$
f(x)\, =\, \ln\left(\sum_{k=1}^K w_k e^{x s_k}\right)\, .
$$
Then $f$ is globally Lipschitz, with Lipschitz constant 1.
\end{lemma}

\noindent
\begin{proof}
Note that
$$
f'(x)\, =\, \frac{\sum_{k=1}^{K} w_k e^{x s_k} s_k}{\sum_{k=1}^K w_k e^{x s_k}}\,
=\, \sum_{k=1}^{K} \theta_k s_k\, ,\qquad \text{where}\quad
\theta_k\, =\, \frac{w_k e^{x s_k}}{\sum_{\ell=1}^{K} w_{\ell} e^{x s_{\ell}}}\quad \text{for}\quad k=1,\dots,K\, .
$$
So $f'(x)$ is in the convex hull of $\{s_1,\dots,s_K\}$, which is a subset of $[-1,1]$ by our assumption.
So $|f'(x)| \leq 1$ for all $x$, which proves the claim.
\end{proof}

By the lemma, we see that $\sfp_N$ is a jointly globally Lipschitz function of the 
random coupling constants $(\sfJ_N(i,j)\, :\, 1\leq i,j\leq N)$. 
Let $\calF_1$ denote the set of all c.d.f.'s $F$ which have finite first moment.
Then, from the above we see that, if $F_N \in \calF_1$, 
then $\sfp_N$ also has a finite first moment.
Therefore, $p_N : \calF_1 \to \R$ is a well-defined function.
Part of our goal is to understand the continuity properties of $p_N$.
Another goal is to derive useful inequalities for the study of ``real''
mean-field spin glass models, such as the SK and VB models.

\subsection{Continuity of $p_N$} 
Given $F \in \calF_1$, 
let us define
$$
a_0(F)\, =\, \int_{-\infty}^{\infty} \ln(\cosh(x))\, dF(x)\, ,
$$
and for $k=1,2,\dots$, define
$$
a_{k}(F)\, =\, \frac{1}{k} \int_{-\infty}^{\infty} \tanh^{k}(x)\, dF(x)\, .
$$

\begin{prop}
\label{prop:A}
Suppose $F_N,\tilde{F}_N \in \calF_1$.
Then
$$
|p_N(F_N) - p_N(\tilde{F})_N|\, \leq\, N\, \sum_{k=0}^{\infty} |a_{k}(F_N) - a_{k}(\tilde{F}_N)|\, .
$$
\end{prop}

\noindent
\begin{proof}
Enumerate all $N^2$ pairs $(i,j)$ in any way, as $(i_1,j_1),\dots,(i_{N^2},j_{N^2})$.
Then the two Hamiltonians associated to $F_N$ and $\tilde{F}_N$ are
$$
-\sfH_N(\s)\, =\, \sum_{n=1}^{N^2} \sfJ_N(i_n,j_n) \s_{i_n} \s_{j_n} + h \sum_{i=1}^N \s_i\qquad \text{and}\qquad
-\tilde{\sfH}_N(\s)\, =\, \sum_{n=1}^{N^2} \tilde{\sfJ}_N(i_n,j_n) \s_{i_n} \s_{j_n} + h \sum_{i=1}^N \s_i\, .
$$
For $n=1,\dots,N^2$ define
$$
\sfw_n(\s)\, =\, \exp\left(\sum_{k=1}^{n-1} \tilde{\sfJ}_N(i_k,j_k) \s_{i_k} \s_{j_k} + \sum_{k=n+1}^{N^2} \sfJ_N(i_k,j_k) \s_{i_k} \s_{j_k}
+ h \sum_{i=1}^N \s_i\right)\, .
$$
Then, by a telescoping sum,
\begin{equation}
\label{eq:ptod}
\begin{split}
\sfp_N - \tilde{\sfp}_N\,
&=\, \frac{1}{N} \sum_{n=1}^{N^2} \left[\ln\left(\sum_{\s \in \Omega_N} \sfw_n(\s) e^{\sfJ_N(i_n,j_n) \s_{i_n} \s_{j_n}}\right)
- \ln\left(\sum_{\s \in \Omega_N} \sfw_n(\s) e^{\tilde{\sfJ}_N(i_n,j_n) \s_{i_n} \s_{j_n}}\right)\right]\\
&=:\, \frac{1}{N} \sum_{n=1}^{N^2} \sfd_n\, .
\end{split}
\end{equation}
Let us define $\hat{\sfw}_n(\s) = \sfw_n(\s) / \sum_{\s \in \Omega_N} \sfw_n(\s)$.
Then we can rewrite
$$
\sfd_n\, =\, \ln\left(\sum_{\s \in \Omega_N} \hat{\sfw}_n(\s)
e^{\sfJ_N(i_n,j_n) \s_{i_n} \s_{j_n}}\right) - \ln\left(\sum_{\s \in
\Omega_N} \hat{\sfw}_n(\s) e^{\tilde{\sfJ}_N(i_n,j_n) \s_{i_n}
\s_{j_n}}\right)\, ,
$$
since the normalizing multipliers cancel in the difference of the logarithms.
Also, let us define
$$
\langle \s_{i_n} \s_{j_n} \rangle_n\, =\, \sum_{\s \in \Omega_N} \hat{\sfw}_n(\s) \s_{i_n} \s_{j_n}\, .
$$
Then, since $e^{x \s} = \cosh(x) + \s \sinh(x)$ for $\pm 1$-valued variables $\s$,
\begin{gather*}
e^{\sfJ_N(i_n,j_n) \s_{i_n} \s_{j_n}}\, =\, \cosh(\sfJ_N(i_n,j_n)) \left[1 + \s_{i_n} \s_{j_n} \tanh(\sfJ_N(i_n,j_n))\right]\\
\Rightarrow\qquad
\ln\left(\sum_{\s \in \Omega_N} \hat{\sfw}_n(\s) e^{\sfJ_N(i_n,j_n) \s_{i_n} \s_{j_n}}\right)\,
=\, \ln[\cosh(\sfJ_N(i_n,j_n))] + \ln\left[1 + \tanh(\sfJ_N(i_n,j_n)) \langle \s_{i_n} \s_{j_n} \rangle_n\right]\, .
\end{gather*}
Since $|\tanh(\sfJ_N(i_n,j_n)) \langle \s_{i_n} \s_{j_n}\rangle_n|<1$, we can use the Taylor expansion of $\ln(1+x)$:
$$
\ln(1+x)\, =\, -\sum_{k=1}^{\infty} \frac{(-1)^k x^k}{k}\, .
$$
Therefore, defining
$$
\phi_0(x)\, =\, \ln(\cosh(x))\qquad \text{and}\qquad \phi_{k}(x)\, =\, \tanh^{k}(x)\qquad \text{for}\qquad k=1,2,\dots\, ,
$$
we have
$$
\sfd_n\, =\, \phi_0(\sfJ_N(i_n,j_n)) - \phi_0(\tilde{\sfJ}_N(i_n,j_n))
- \sum_{k=1}^{\infty} \frac{(-1)^{k} \left( \langle \s_{i_n} \s_{j_n}\rangle_n \right)^k}{k} \left[\phi_{k}(\sfJ_N(i_n,j_n))
- \phi_{k}(\tilde{\sfJ}_N(i_n,j_n))\right]\, .
$$
Importantly,  by our assumption, $\sfJ_N(i_n,j_n)$ and $\tilde{\sfJ}_N(i_n,j_n)$ are independent of 
all $\sfJ_N(i_k,j_k)$ aand $\tilde{\sfJ}_N(i_k,j_k)$ for $k\neq n$.
Therefore, $\sfJ_N(i_n,j_n)$ and $\tilde{\sfJ}_N(i_n,j_n)$ are both independent of 
$\langle \s_{i_n} \s_{j_n}\rangle_n$.
Also, obviously,
$\E[\phi_{k}(\sfJ)] = a_{k}(F)$ for an $F$-distributed random variable $\sfJ$.
$$
\E[\sfd_n]\, =\, a_0(F) - a_0(\tilde{F}) - \sum_{k=1}^{\infty} (-1)^k\, \E\left[ \left( \langle \s_{i_n} \s_{j_n}\rangle_n \right)^k\right] [a_k(F) - a_k(\tilde{F})]\, .
$$
Since, $|\langle \s_{i_n} \s_{j_n}\rangle_n|<1$, 
$$
|\E[\sfd_n]|\, \leq\, \sum_{k=0}^{\infty} |a_{k}(F_N) - a_{k}(\tilde{F}_N)|\, .
$$
This is true for all $1\leq n\leq N^2$.
Plugging into (\ref{eq:ptod}), we obtain the result.
\end{proof}

\subsection{Re-derivation of a result of Carmona and Hu}

In an important paper, Carmona and Hu proved universality for the Sherrington-Kirkpatrick model \cite{CarmonaHu}.
Using the proposition, we can re-derive part of their results, in fact with slightly weaker hypotheses.
Suppose $F$ is a cumulative distribution function such that 
\begin{equation}
\label{eq:cond1}
\mu_2(F)\, :=\, \int_{-\infty}^{\infty} x^2\, dF(x)\, =\, \beta^2/2\, ,
\end{equation}
and
\begin{equation}
\label{eq:cond2}
\mu_1(F)\, :=\, \int_{-\infty}^{\infty} x\, dF(x)\, =\, 0\, .
\end{equation}
In other words, letting $\sfJ$ be distributed according to $F$, we have $\E[\sfJ]=0$ and ${\rm Var}(\sfJ)=\beta^2/2$.
The c.d.f.\ of $\sfJ/\sqrt{N}$, then, is $F_N(x)\, =\, F(\sqrt{N} x)$.
Let us write $\sfp_N(F_N)$ just to make explicit
the dependence of $\sfp_N$ on the distribution of all the random coupling constants $(\sfJ_N(i,j)\, :\, 1\leq i,j\leq N)$.

Let $F^*_{\beta}$ be the c.d.f.\ for an $\mathcal{N}(0,\beta^2/2)$ random variable and $F^*_{N,\beta}(x) = F^*_{\beta}(\sqrt{N} x)$.
Then, defining $p_N^{\rm SK}(\beta) := p_N(F^*_{N,\beta})$, this is the quenched pressure for the SK model at inverse temperature $\beta$.
By an important result of Guerra and Toninelli, the thermodynamic limit, $p^{\rm SK}(\beta) := \lim_{N \to \infty} p_N^{\rm SK}(\beta)$, exists.
Among other important results, Carmona and Hu proved the following result, however with slightly different hypotheses:
\begin{cor}
\label{cor:CarmonaHu}
Suppose $F$ satisfies (\ref{eq:cond1}) and (\ref{eq:cond2}), and $F_N(x) = F(\sqrt{N} x)$ for all $N>0$.
Then
$$
\sfp_N(F_N) \to p^{\rm SK}(\beta)\, ,
$$
as $N \to \infty$,
almost surely, and in probability.
\end{cor}
This is a non-quantitative version of Carmona and Hu's Theorem 1.
They obtained explicit bounds on $|p_N(F_N) - p^{\rm SK}{N}(\beta)|$.
On the other hand, to do so, they assumed that $F$ also has a finite third moment, which is slightly
stronger than our hypotheses.
(But without such an assumption, it is impossible to get quantitative bounds.)
We will re-prove Corollary \ref{cor:CarmonaHu} in order to demonstrate that
it can be derived from Proposition \ref{prop:A}, rather than from Carmona and Hu's alternative,
and very interesting, approach of ``approximate Gaussian integration by parts''.
The reason we do this is that later we will use Proposition \ref{prop:A} to
prove an analogous result for the Viana-Bray model.
(Incidentally, Carmona and Hu's result was extended
to quantum spin glasses by Crawford in \cite{Crawford}.
We have not considered whether a similar generalization
is possible for Proposition \ref{prop:A}.)

We can prove convergence in mean, easily, using the proposition.
But then convergence in probability follows trivially using an important
and well-known result of Pastur and Shcherbina.
In an important paper, they proved that the spin-spin overlap order parameter
of the SK model is not self-averaging in a certain regime \cite{PasturShcherbina}.
They also proved the much simpler, but even more widely cited, result
in an appendix, that the random pressure is self-averaging.
It is that second result which we now state, and for completeness prove, in the context
of general mean-field spin glasses.

\begin{lemma}
\label{lem:PasturShcherbina}
Let $F_N$ satisfy $\mu_2(F_N)<\infty$.
Let $\sfp_N$ denote the random pressure associated to the distribution
of the coupling constants with c.d.f.\ $F_N$.
Then ${\rm Var}(\sfp_N) \leq \s^2(F_N) = \mu_2(F_N) - \mu_1(F_N)^2$.
\end{lemma}

\noindent
\begin{proofof}{\bf Proof of Lemma \ref{lem:PasturShcherbina}:}
The proof follows the standard martingale method.
Again, enumerate all $N^2$ pairs $(i,j)$ as $(i_1,j_1),\dots,(i_{N^2},j_{N^2})$.
Let $\calF_n = \s(\sfJ(i_k,j_k)\, :\, k\leq n)$ be the $\s$-algebra generated by $\sfJ(i_k,j_k)$
for $k\leq n$.
Let $\calF_0$ be the trivial $\s$-algebra.
For $n=0,1,\dots,N^2$, define $\sfM_n\, =\, \E[\sfp_N\, |\, \calF_n]$.
Note that $\sfp_N = \sfM_{N^2}$ while $\E[\sfp_N] = \sfM_0$.
Therefore, by a telescoping sum,
\begin{equation}
\label{eq:mart}
\Var(\sfp_N)\, =\, \sum_{n=1}^{N^2} \E[\sfM_n^2-\sfM_{n-1}^2]\, .
\end{equation}
In addition to the coupling constants $\sfJ_N(i,j)$ distributed according to $F_N$,
let us define additional independent copies of
these random variables, called $\sfJ^{1}_N(i,j)$ and $\sfJ^{2}_N(i,j)$.
Define, for $\alpha,\beta \in \{1,2\}$,
$$
-\sfH_{N,n}^{\alpha,\beta}(\s)\, =\, \sum_{k=1}^{n-1} \sfJ_N(i_k,j_k) \s_{i_k} \s_{j_k}
+ \sfJ^{\alpha}_N(i_n,j_n) \s_{i_n} \s_{j_n}
+  \sum_{k=n+1}^{N^2} \sfJ_N^{\beta}(i_k,j_k) \s_{i_k} \s_{j_k}
+ h \sum_{i=1}^N \s_i\, .
$$
Also define
$$
\sfp_{N,n}^{\alpha,\beta}\, =\, \frac{1}{N} \ln\left(\sum_{\s \in \Omega_N} \exp(-\sfH_{N,n}^{\alpha,\beta}(\s))\right)\, .
$$
Then
$$
\E[\sfM_n^2-\sfM_{n-1}^2]\, =\, \frac{1}{2} \E[(\sfp_{N,n}^{1,1} - \sfp_{N,n}^{2,1})(\sfp_{N,n}^{1,2}-\sfp_{N,n}^{2,2})]\, .
$$
Finally, we want to bound this.
So, define
$$
-\sfH_{N,n}^{\emptyset,\beta}(\s)\, =\, \sum_{k=1}^{n-1} \sfJ_N(i_k,j_k) \s_{i_k} \s_{j_k}
+  \sum_{k=n+1}^{N^2} \sfJ^{\beta}_N(i_k,j_k) \s_{i_k} \s_{j_k}
+ h \sum_{i=1}^N \s_i\, .
$$
Then, for $\beta=1,2$,
$$
|\sfp_{N,n}^{1,\beta} - \sfp_{N,n}^{2,\beta}|\,
=\, \frac{1}{N} \left| \ln\left(\sum_{\s \in \Omega_N} e^{-\sfH_{N,n}^{\emptyset,\beta}(\s)} e^{\sfJ^{1}_N(i_n,j_n) \s_{i_n} \s_{j_n}}\right)
-  \ln\left(\sum_{\s \in \Omega_N} e^{-\sfH_{N,n}^{\emptyset,\beta}(\s)} e^{\sfJ^{2}_N(i_n,j_n) \s_{i_n} \s_{j_n}}\right)\right|\, .
$$
By Lemma \ref{lem:basic}, the right hand side is bounded above by $N^{-1} |\sfJ^{1}_N(i_n,j_n) - \sfJ^{2}_N(i_n,j_n)|$.
Therefore,
$$
\E[\sfM_n^2-\sfM_{n-1}^2]\, \leq\, \frac{1}{2N^2} \E[|\sfJ_N^{1}(i_n,j_n) - \sfJ_N^{2}(i_n,j_n)|^2]\, =\, \frac{\s^2(F_N)}{N^2}\, .
$$
Using this bound for all $1\leq n\leq N^2$, and plugging in to (\ref{eq:mart}) gives the result.
\end{proofof}

\noindent
\begin{proofof}{\bf Proof of Corollary \ref{cor:CarmonaHu}:}
By the triangle inequality
and Proposition \ref{prop:A}, we can bound
$$
|p_N(F_N) - p^{\rm SK}_N(\beta)|\, 
\leq\, \delta_N(F_N) + \delta_N(F_{N,\beta}^*)\, ,
$$
where
$$
\delta_N(F_N)\, =\,
\left|N a_0(F_N) - \frac{\beta^2}{4}\right| 
+  \left|N a_2(F_N) - \frac{\beta^2}{4}\right|
+ N \sum_{\substack{k=1 \\ k \neq 2}}^{\infty} |a_k(F_N)|\, .
$$
We will show that $\delta_N(F_N) \to 0$, as $N \to \infty$.
Since $F_\beta^*$ also satisfies (\ref{eq:cond1}) and (\ref{eq:cond2}),
this also implies $\delta_N(F_{N,\beta}^*) \to 0$ as $N \to \infty$.
Note that
\begin{align*}
\frac{\beta^2}{4} - N a_0(F_N)\, 
&=\, \int_{-\infty}^{\infty} \frac{x^2}{2}\, dF(x) - \int_{-\infty}^{\infty} N \ln(\cosh(x))\, dF_N(x)\\
&=\, \int_{-\infty}^{\infty} \left(\frac{x^2}{2} - N \ln\left(\cosh\left(\frac{x}{\sqrt{N}}\right)\right)\right)\, dF(x)\, .
\end{align*}
Note that, since $\ln(\cosh(x)) \leq x^2/2$, the integrand is nonnegative. Similarly,
$$
\frac{\beta^2}{4} - N a_2(F_N)\,
=\, \int_{-\infty}^{\infty} \left(\frac{x^2}{2} - \frac{N}{2} \tanh^2\left(\frac{x}{\sqrt{N}}\right)\right)\, dF(x)\, ,
$$
and the integrand is nonnegative because $\tanh^2(x) \leq x^2$.
Also, notice that $a_{2k}(F_N) \geq 0$ for all $k=0,1,\dots$, because $\tanh^{2k}(x) \geq 0$,
and also
$$
\sum_{k=2}^{\infty} \frac{\tanh^{2k}(x)}{2k}\,
=\, - \frac{1}{2} \ln(1 - \tanh^2(x)) - \tanh^2(x)\, =\, \ln(\cosh(x)) - \frac{\tanh^2(x)}{2}\, .
$$
Therefore,
$$
\left|N a_0(F_N) - \frac{\beta^2}{4}\right| 
+  \left|N a_2(F_N) - \frac{\beta^2}{4}\right|
+ N \sum_{k=2}^{\infty} |a_{2k}(F_N)|\, 
=\, \int_{-\infty}^{\infty} \left( x^2 - N \tanh^2\left(x/\sqrt{N}\right) \right)\, dF(x)\, .
$$
By the dominated convergence theorem, the last quantity approaches $0$ as $N\to \infty$.
Note that, since $\mu_1(F)=0$,
\begin{align*}
-N a_1(F_N)\, =\, N(\mu_1(F_N) - a_1(F_N))\,
&=\, N \int_{-\infty}^{\infty} (x - \tanh(x))\, dF_N(x)\\ 
&=\, N \int_{-\infty}^{\infty} \left( \int_0^x \tanh^2(y)\, dy\right)\, dF_N(x)\\ 
&=\, N \int_{0}^{\infty} \tanh^2(y)\, \int\nolimits_{|x| \geq \sqrt{N}\, y} {\rm sgn}(x)\, dF(x)\, dy\, .
\end{align*}
So,
$$
N |a_1(F_N)|\, \leq\, \int_{0}^{\infty} \frac{\tanh^2(y)}{y^2}\, \int\nolimits_{|x| \geq \sqrt{N}\, y} x^2\, dF(x)\, dy\, .
$$
Since $\mu_2(F) < \infty$, this quantity also approaches $0$ as $N \to \infty$, by the dominated convergence theorem.
Finally,
\begin{align*}
\sum_{k=1}^{\infty} |a_{2k+1}(F_N)|\,
&=\, \sum_{k=1}^{\infty} \frac{1}{2k+1} \left|\int_{-\infty}^{\infty} \tanh^{2k+1}(x)\, dF_N(x)\right|\\
&\leq\, \int_{-\infty}^{\infty} \sum_{k=1}^{\infty} \frac{|\tanh^{2k+1}(x)|}{2k+1}\, dF_N(x)\\
&=\, \int_{-\infty}^{\infty} \left( \frac{1}{2} \ln\left(\frac{1+|\tanh(x)|}{1-|\tanh(x)|}\right) - |\tanh(x)| \right)\, dF_N(x)\\
&=\, \int_{-\infty}^{\infty} (|x| - |\tanh(x)|)\, dF_N(x)\, .
\end{align*}
Therefore, the same argument as the one just above shows that
$$
\lim_{N \to \infty} N \sum_{k=1}^{\infty} |a_{2k+1}(F_N)|\, =\, 0\, .
$$ 
This completes the proof that $\lim_{N \to \infty} \delta_N(F_N) = 0$.
Therefore, it also shows that $\lim_{N \to \infty} |p_N(F_N) - p^{\rm SK}_N(\beta)| = 0$.
Since $\lim_{N \to \infty} p^{\rm SK}_N(\beta) = p^{\rm SK}(\beta)$, in order to complete the proof,
all we need to show is that $\sfp_N(F_N) - p_N(F_N) \to 0$, as $N \to \infty$, in probability.
But, by Pastur and Shcherbina's bound,
$$
\E[(\sfp_N(F_N) - p_N(F_N))^2]\, \leq\, \mu_2(F_N)\, =\, \frac{\mu_2(F)}{N}\, .
$$
So, we have the even stronger result: $\sfp_N(F_N) - p_N(F_N) \to 0$ in $L^2$.
\end{proofof}

\subsection{An application to the Viana-Bray model}
Now, let us now reconsider Proposition \ref{prop:A} in the context of the Viana-Bray model.
Define $p^{{\rm VB}}_N(\alpha,\beta) = p_N(F^*_{N,\alpha,\beta})$,
where
$$
F^*_{N,\alpha,\beta}\, =\, e^{-\alpha/N} \sum_{k=0}^{\infty} \frac{(\alpha/N)^k}{k!} (F^*_{1,\beta})^{\star k}\, ,
$$
in which $\star$ is the convolution product.
Hence, defining $\sfK$ to be a Poisson-$(\alpha/N)$ random variable,
and defining $\sfg_1,\sfg_2,\dots$ to be i.i.d., $\mathcal{N}(0,1)$ random variables,
$F^*_{N,\alpha,\beta}$ is the c.d.f.\ for the random variable
$$
\sfJ\, =\, \sum_{k=1}^{\sfK} \sfg_k\, .
$$
This is the ``Poissonized Gaussian'' coupling used in one version of the Viana-Bray model.

\begin{cor}
\label{cor:VB1}
Suppose that, for each $N>0$, the sequence $(f_{N,0},f_{N,1},\dots)$
is a probability mass function, such that
$$
\lim_{N \to \infty} N f_{N,1}\, =\, \alpha\qquad \text{and}\qquad
\lim_{N \to \infty} N \sum_{k=2}^{\infty} k f_{N,k} \, =\, 0\, .
$$
For each $N>0$, define
$$
F_N\, =\, \sum_{k=0}^{\infty} f_{N,k}\, (F^*_{1,\beta})^{\star k}\, .
$$
Then
$$
\lim_{N \to \infty} |p_N(F_N) - p_N^{{\rm VB}}(\alpha,\beta)|\, =\, 0\, .
$$
\end{cor}

\noindent
As a particular application, we could take $f_{N,k} = (1-\frac{\alpha}{N}) \delta_{k,0} + \frac{\alpha}{N} \delta_{k,1}$.
I.e., instead of taking $\sfK$ to be Poisson, with mean $\alpha/N$, we could take it to be Bernoulli
with the same mean.

\smallskip
\noindent
\begin{proof}
Note that
\begin{align*}
\sum_{k=0}^{\infty} |a_{2k}(F_N) - f_{N,1} a_{2k}(F^*_{1,\beta})|\,
&=\, \int_{-\infty}^{\infty} 2 \ln(\cosh(x))\, [dF_N(x) - f_{N,1}\, dF^{*}_{1,\beta}(x)]\\
&\leq\, \int_{-\infty}^{\infty} x^2\, \sum_{k=2}^{\infty} f_{N,k}\, d (F^{*}_{1,\beta})^{\star k}(x)\\
&=\, \frac{\beta^2}{2} \sum_{k=2}^{\infty} k\, f_{N,k}\, .
\end{align*}
Defining $f^*_{N,k}(\alpha) = e^{-\alpha/N} (\alpha/N)^k/k!$, we see that exactly the same property is true
of it.
But also, by assumption and calculation,
$$
\lim_{N \to \infty} N\, \sum_{k=2}^{\infty} k\, f^*_{N,k}(\alpha)\,
=\, \lim_{N \to \infty} \alpha (1-e^{-\alpha/N})\,
=\, 0\,
=\, \lim_{N \to \infty} N\, \sum_{k=2}^{\infty} k\, f_{N,k}\, .
$$
Therefore,
\begin{align*}
\limsup_{N \to \infty} N \sum_{k=0}^{\infty} |a_{2k}(F_N) - a_{2k}(F^*_{N,\alpha,\beta})|\,
&=\, \limsup_{N \to \infty} N |f_{N,1} - f^*_{N,1}(\alpha)| \sum_{k=0}^{\infty} a_{2k}(F^*_{1,\beta})\\
&\leq\, \frac{\beta^2}{2} \limsup_{N \to \infty} N |f_{N,1} - f^*_{N,1}(\alpha)|\, .
\end{align*}
But, of course, by assumption and calculation
$$
\lim_{N \to \infty} N f^*_{N,1}(\alpha)\, =\, \lim_{N \to \infty} e^{-\alpha/N} \alpha\, =\, \alpha\, =\, \lim_{N \to \infty} N\, f_{N,1}\, .
$$
Since $a_k(F_N) = a_k(F^*_{N,\alpha,\beta})=0$, for odd $k$, because the Gaussian is symmetric,
we then see that
$$
\lim_{N \to \infty} N \sum_{k=0}^{\infty} |a_{k}(F_N) - a_{k}(F^*_{N,\alpha,\beta})|\, =\, 0\, ,
$$
and we can apply Proposition \ref{prop:A}.
\end{proof}

\noindent
Also note that, again, Pastur and Shcherbina's self-averaging bounds prove that $\sfp_N - p_N(F_N) \to 0$,
in $L^2$, as $N \to \infty$.
Indeed,
$$
\s^2(F_N)\, =\, \sum_{k=0}^{\infty} f_{N,k} \s^2((F^*_{1,\beta})^{\star k})\, =\, \frac{\beta^2}{2} \sum_{k=0}^{\infty} k f_{N,k}\, ,
$$
and the right-hand-side goes to $0$, as $N \to \infty$, by hypothesis.

\subsubsection{Poisson thinning.}

Let us briefly address one possible point of confusion. The Hamiltonian we wrote for the Viana-Bray model was
$$
-\sfH_N(\s)\, =\, \sum_{i,j=1}^N \sum_{k=1}^{\sfK(i,j)} \sfg_k(i,j) \s_i \s_j + h \sum_{i=1}^N \s_i\, ,
$$
where, for each $(i,j) \in \{1,\dots,N\} \times \{1,\dots,N\}$ the random variable $\sfK(i,j)$ is a Poisson random variable, with mean $\alpha/N$,
such that all the variables $\{\sfK(i,j)\, :\, 1\leq i,j\leq N\}$ are independent,
and all the random variables $\sfg_{k}(i,j)$, for $k=1,2,\dots$,
are i.i.d.\ $\mathcal{N}(0,\beta^2)$ random variables, all of which are independent, and independent of the $\sfK(i,j)$'s.
This is not literally the model that is written down in some references on the Viana-Bray model.
Let us call the model we wrote above the {\it Poissonized Viana Bray model}\footnote{We thank an
anonymous referee for suggesting this name, as well as for raising the issue of demonstrating
the fact that the two versions of the model are statistically equivalent.}.
The original Viana-Bray model considered by many authors is 
$$
-\tilde{\sfH}_N(\s)\, =\, \sum_{k=1}^{\boldsymbol{\sfK}} \sfg_{k}\, \s_{\sfi_k} \s_{\sfj_k} + h \sum_{i=1}^{N} \s_i\, ,
$$
where $\boldsymbol{\sfK}$ is a Poisson random variable, with mean $\alpha N$, and $\sfi_1,\sfi_2,\dots$ and $\sfj_1,\sfj_2,\dots$
are i.i.d., random variables, uniformly distributed on the $N$ sites $\{1,\dots,N\}$.
Also, $\sfg_1,\sfg_2,\dots$ are i.i.d., $\mathcal{N}(0,\beta^2)$ random variables, independent of everything else.

These two versions of the model are statistically equivalent.
So all expectations of all functions of the Hamiltonians
are equal, including the quenched pressure.
To see this, we use a well-known property of Poisson random variables, which is commonly 
called ``Poisson thinning''.
We can construct a direct correspondence between the random variables of the first Hamiltonian
and the second one.
For instance, given $\boldsymbol{\sfK}$, $\sfi_1,\sfi_2,\dots$, $\sfj_1,\sfj_2,\dots$ and $\sfg_1,\sfg_2,\dots$, do the following:
First, let $\tilde{\sfK}(i,j) = \#\{k\, :\, k \leq \boldsymbol{\sfK}\, ,\ \sfi_k=i\, ,\ \sfj_k=j\}$ for each $1\leq i,j\leq N$.
Note that
\begin{align*}
\E\left[\exp\left(\sum_{i,j=1}^{N} \lambda_{i,j} \tilde{\sfK}(i,j)\right)\right]\,
&=\, \E\left[\exp\left(\sum_{k=1}^{\boldsymbol{\sfK}} \lambda_{\sfi_k,\sfj_k}\right)\right]\,
=\, \E\left[\left(N^{-2} \sum_{i,j=1}^{N} e^{\lambda_{i,j}}\right)^{\boldsymbol{\sfK}}\right]\\
&=\, \exp\left(\alpha N \left(N^{-2} \sum_{i=1}^{N} \sum_{j=1}^{N} e^{\lambda_{i,j}} - 1\right)\right)\,
=\, \prod_{i,j=1}^{N} \exp\left(\frac{\alpha}{N} [e^{\lambda_{i,j}} - 1]\right)\, ,
\end{align*}
which is exactly the joint moment generating function for i.i.d., Poisson random variables with mean $\alpha/N$.
Therefore, the random variables $\tilde{\sfK}(i,j)$ have identical joint distribution to $\sfK(i,j)$.
Similarly, there is a way to construct $\tilde{\sfg}_k(i,j)$'s from the $\sfg_k$'s, by merely letting 
$\tilde{\sfg}_k(i,j)$ equal $\sfg_{\sfn_k(i,j)}$ where $\sfn_k(i,j)$ is the $k$th smallest integer $n$ such that 
$(\sfi_n,\sfj_n)=(i,j)$.
Using independence, it is trivial to see that the collection of $\{\tilde{\sfg}_k(i,j)\, :\, 1\leq i,j\leq N\, ,\ k=1,2,\dots\}$
is equivalent to the collection of $\{\sfg_k(i,j)\, :\, 1\leq i,j\leq N\, ,\ k=1,2,\dots\}$.

\section{Continuity for non-independent couplings}

A key ingredient in the proof of Proposition \ref{prop:A} was the assumption that all the coefficients were independent.
But for some purposes, one wants to drop that assumption, instead assuming that all the coefficients
$\sfJ(i,j)$ and $\tilde{\sfJ}(i,j)$ are defined on a common probability space,
and are close in some sense.
Let us state a bound that works in that case.

\begin{prop}
\label{prop:B}
Suppose the following random variables are defined on one probability space:
a random integer $\sfN \geq 0$;
random spin sites $\sfi_1,\dots,\sfi_{\sfN},\sfj_1,\dots,\sfj_{\sfN} \in \{1,\dots,N\}$;
and random couplings $\sfJ_1,\dots,\sfJ_{\sfN}$ and $\tilde{\sfJ}_1,\dots,\tilde{\sfJ}_N$,
which may be dependent and possibly not identically distributed.
Define the random Hamiltonians
$$
-\sfH_N(\s)\, =\, \sum_{n=1}^{\sfN} \sfJ_n \s_{\sfi_n} \s_{\sfj_n} + h \sum_{i=1}^N \s_i\qquad \text{and}\qquad
-\tilde{\sfH}_N(\s)\, =\, \sum_{n=1}^{\sfN} \tilde{\sfJ}_n \s_{\sfi_n} \s_{\sfj_n} + h \sum_{i=1}^N \s_i\, .
$$
Then
$$
\left|\frac{1}{N}\, \E\left[\ln\left( \sum_{\s \in \Omega_N} e^{-\sfH_N(\s)}\right)\right]
- \frac{1}{N}\, \E\left[\ln\left( \sum_{\s \in \Omega_N} e^{-\tilde{\sfH}_N(\s)}\right)\right]\right|\,
\leq\, \frac{1}{N}\, \E\left[\sum_{n=1}^{\sfN} |\sfJ_n - \tilde{\sfJ}_n|\right]\, .
$$
\end{prop}

\noindent
\begin{proof}
Consider the linear interpolation
$\sfH_{N,t}=t\sfH_N+(1-t)\tilde{\sfH}_N$.
Then
$$
\frac{d}{dt} \E\left[\ln\left(\sum_{\s \in \Omega_N} e^{-\sfH_{N,t}(\s)}\right)\right]\,
=\, - \E\left[\frac{\sum_{\s \in \Omega_N} e^{-\sfH_{N,t}(\s)} \left(\sfH_N(\s)-\tilde{\sfH}_N(\s)\right)}
{\sum_{\s \in \Omega_N} e^{-\sfH_{N,t}(\s)}}\right]\, ,
$$
and integrating
over $(0,1)$ it readily follows that
$$
\left|\frac{1}{N}\, \E\left[\ln\left( \sum_{\s \in \Omega_N} e^{-\sfH_N(\s)}\right)\right]
- \frac{1}{N}\, \E\left[\ln\left( \sum_{\s \in \Omega_N} e^{-\tilde{\sfH}_N(\s)}\right)\right]\right|\,
\leq\, \frac{1}{N}
\E\left[\max_{\s\in
\Omega_N}\left|\sfH_N(\s)-\tilde{\sfH}_N(\s)\right|\right]\, .
$$
Now conditioned on $\sfN$, for all $\s\in\Omega_N$,
$$
\left|\sfH_N(\s)-\tilde{\sfH}_N(\s)\right|\,
\leq\,
\sum_{n=1}^{\sfN} |\sfJ_{n} - \tilde{\sfJ}_n|\, .
$$
Taking expectations gives the desired result.
\end{proof}

\subsection{A canonical, versus grand canonical, version of the Viana-Bray model}
An equivalent definition of the Hamiltonian for the
Viana-Bray model \cite{VB} is
$$
-\sfH_N(\s)\, =\, \sum_{n=1}^{\sfK}\sfJ_{n}\, \s_{\sfi_n}
\s_{\sfj_n} + h \sum_{i=1}^{N} \s_i\, ,
$$
where: $\sfK$ is a Poisson random variable, with mean $\alpha N$;
the $\sfJ_{n}$'s are
i.i.d., $\mathcal{N}(0,\beta^2)$ Gaussian random variables;
and $\sfi_1,\sfi_2,\dots,\sfj_1,\sfj_2,\dots$ are i.i.d.,
integer-valued random variables, uniformly distributed on
the set $\{1, 2, \dots, N\}$.
Then,
$$
p_N^{\rm VB}(\alpha,\beta)\, =\, \frac{1}{N}\, \E\left[\ln\left( \sum_{\s \in \Omega_N} e^{-\sfH_N(\s)}\right)\right]\, .
$$
Note that the number of edges present is a random variable $\sfK$.

In statistical mechanics, we often consider the canonical ensemble to be a model of a gas
where the number of particles (but not the energy) is held fixed.
In the grand canonical ensemble, the number of particles is random, although usually
highly concentrated around its mean value.
Since the number of edges in the Viana-Bray model is random (but highly concentrated
around its average), this is like a grand canonical ensemble for the number of edges.
Consider an alternative ``canonical ensemble'' Hamiltonian
$$
-\sfH^{\rm can}_N(\s)\,
=\, \sum_{n=1}^{\lfloor \alpha N \rfloor}\sfJ_{n}\, \s_{\sfi_n} \s_{\sfj_n} + h \sum_{i=1}^{N} \s_i\, ,
$$
where all the $\sfJ_n$'s are i.i.d., $\mathcal{N}(0,\beta^2)$ Gaussian random variables,
and $\sfi_1,\sfi_2,\dots,\sfj_1,\sfj_2,\dots$ are as before.
Now the number of edges is nonrandom: it is $\lfloor \alpha N \rfloor$ is nonrandom, the greatest integer $\leq \alpha N$.
Define
$$
p^{\rm can}_N(\alpha,\beta)\, =\,  \frac{1}{N}\, \E\left[\ln\left( \sum_{\s \in \Omega_N} e^{-\tilde{\sfH}_N(\s)}\right)\right]\, .
$$

\begin{cor}
With the definitions above,
$$
p^{\rm can}_N(\alpha,\beta) - p_N^{\rm VB}(\alpha,\beta)\, =\, O\left(\frac{1}{\sqrt{N}}\right)\, .
$$
\end{cor}

\begin{remark}
A qualitative version of this result was stated without proof in the papers of Franz and Leone \cite{FranzLeone}
and Guerra and Toninelli \cite{GT3}.
\end{remark}

\noindent
\begin{proof}
Define $\sfg_1,\sfg_2,\dots$ to be i.i.d.\ $\mathcal{N}(0,1)$ random variables.
Let $\sfK$ be a $\mathcal{P}(\alpha N)$ random variable, and let $\sfi_1,\sfi_2,\dots,\sfj_1,\sfj_2,\dots$
be i.i.d., and uniform on $\{1,\dots,N\}$, as before.
For $n\leq \sfK$, let $\sfJ_n = \beta \sfg_n$.
For $n> \sfK$, define $\sfJ_n = 0$.
For $n\leq \lfloor \alpha N \rfloor$ define $\tilde{\sfJ}_n = \beta \sfg_n$.
For $n >  \lfloor \alpha N \rfloor$ define $\tilde{\sfJ}_n = 0$.
Let $\sfN = \max(\sfK,\lfloor \alpha N \rfloor)$.
It is easy to see that $\sfH_N(\s)$ and $\tilde{\sfH}_N(\s)$, defined as in Prop \ref{prop:B},
have the correct distributions for the Viana-Bray and the ``canonical ensemble'' models, respectively.
Therefore,
\begin{align*}
|p^{\rm can}_N(\alpha,\beta) - p_N^{\rm VB}(\alpha,\beta)|\,
\leq\, \frac{1}{N}\, \E\left[\sum_{n=1}^{\sfN} |\sfJ_n - \tilde{\sfJ}_n|\right]\,
&=\, \beta\, \frac{1}{N}\, \E[|\sfK - \lfloor \alpha N \rfloor|]\, \E[|\sfg_N|]\\
&\leq\, \frac{\beta}{N} (1 + \operatorname{Var}(\sfK)^{1/2})\, \operatorname{Var}(\sfg_N)^{1/2}\\
&=\, \frac{\beta}{N} (1 + \sqrt{\alpha N})\, .
\end{align*}
\end{proof}

\subsection{The canonical, versus grand canonical, version of the SK model}
An analogous result holds for the SK model.
A definition of the SK model is as follows:
Let $\sfX_N$ be a $\chi_{N^2}$ random variable.
Let $\sfV_N = (\sfV(i,j)\, :\, 1\leq i,j\leq N)$ be a uniform random point on the unit sphere
$$
\mathbb{S}^{N^2-1}\, =\, \Big\{V = (V(i,j)\, :\, 1\leq i,j\leq N)\, \Big|\, \sum\nolimits_{i,j=1}^N (V(i,j))^2\, =\, 1\Big\}\, .
$$
Then, defining $\sfJ_N(i,j) = \beta \sfV_N(i,j) \sfX_N/\sqrt{2N}$,
$$
p^{\rm SK}_N(\beta)\, =\, \frac{1}{N}\, \E\left[\ln\left(\sum_{\s \in\Omega_N} \exp\left(\sum_{i,j=1}^{N} \sfJ(i,j) \s_i \s_j + h \sum_{i=1}^N \s_i\right)\right)\right]\, .
$$
Again, note that the norm of the coupling constant vector, $\vec{\sfJ}_N = (\sfJ_N(i,j)\, :\, 1\leq i,j\leq N)$, equals $\sfX_N/\sqrt{2N}$,
which is random itself.
One could consider this to be a grand canonical ensemble.
In the ``canonical ensemble'' the only thing random about the coupling constant vector would be the direction.
Therefore, define $\tilde{\sfJ}_N(i,j)\, =\, \beta \sfV(i,j) \sqrt{N/2}$, and define
$$
p^{\rm can}_N(\beta)\, =\, \frac{1}{N} \E\left[\ln\left(\sum_{\s \in\Omega_N} \exp\left(\sum_{i,j=1}^{N} \tilde{\sfJ}(i,j) \s_i \s_j + h \sum_{i=1}^N \s_i\right)\right)\right]\, .
$$
This is the analogue of the SK model, but where the couplings constant vector, $(\tilde{\sfJ}_N(i,j)\, :\, 1\leq i,j\leq N)$, is
constrained to lie on a sphere with radius $R_N$ and 
satisfying $R_N^2 = \beta^2 N/2 = \E[\|\vec{\sfJ_N}\|^2]$.

\begin{cor}
With the definitions above,
$$
p^{\rm can}_N(\beta) - p_N^{\rm SK}(\beta)\, =\, O\left(\frac{1}{\sqrt{N}}\right)\, .
$$
\end{cor}

\noindent
\begin{proof}
By direct application of Prop \ref{prop:B},
\begin{align*}
|p^{\rm can}_N(\beta) - p_N^{\rm SK}(\beta)|\,
&\leq\, \frac{1}{N}\, \E\left[\sum_{i,j=1}^{N} |\sfJ_N(i,j) - \tilde{\sfJ}_N(i,j)|\right]\\
&=\, \frac{\beta}{\sqrt{2N^3}}\, \sum_{i,j=1}^N \E[|\sfV_N(i,j)|\cdot |\sfX_N - N|]\\
&\leq\,  \frac{\beta}{\sqrt{2N^3}}\, \sum_{i,j=1}^N (\E[\sfV_N(i,j)^2])^{1/2}\, \left(\E[(\sfX_N-N)^{2}]\right)^{1/2}\\
&=\, \frac{\beta}{\sqrt{2N}}\, \left(\E[\sfX_N^2] + N^2 - 2 N\E[\sfX_N]\right)^{1/2}\, .
\end{align*}
But, as $\sfX_N$ is $\chi_{N^2}$, that means $\E[\sfX_N^2] = N^2$, while
$$
\E[\sfX_N]\, =\, \frac{\sqrt{2}\, \Gamma((N^2+1)/2)}{\Gamma(N^2/2)}\, =\, N - O\left(\frac{1}{N}\right)\, .
$$
Therefore,
$$
\E[\sfX_N^2] + N^2 - 2 N\E[\sfX_N]\, =\, O(1)\, .
$$
\end{proof}

\section{Mean-field spin glass models with infinitely divisible couplings}

All of the results so far were motivated by the beautiful results for the Viana-Bray
models obtained by Franz and Leone \cite{FranzLeone}, Guerra and Toninelli \cite{GT3},
and De Sanctis \cite{DeSanctis}.
Additionally, we were motivated by Carmona and Hu's universality result for the SK model \cite{CarmonaHu}.
The Viana-Bray model, as studied in the papers listed above, basically relies upon one property
of the coupling distribution.
That is that $F_N$ is infinitely divisible.
Let us digress briefly, to discuss infinitely divisible distributions.

We will specialize our attention to symmetric distributions.
Suppose that $\Lambda$ is a nonnegative measure on $(0,\infty)$
(not including $0$) satisfying
$$
\int_{0+}^{\infty} \min(y^2,1)\, \Lambda(dy)\, <\, \infty\, .
$$
(We write $\int_{0+}^{\infty}$ in place of $\int_{(0,\infty)}$.)
Also suppose $v$ is a nonnegative number.
Then one can define a function
$$
\Psi_{(\Lambda,v)}(k)\, =\, \frac{v k^2}{2} + \int_{0+}^{\infty} (1 - \cos(ky))\, \Lambda(dy)\, .
$$
This function is conditionally negative semidefinite, and $0$ at $0$.
In other words (c.f.\ Schoenberg's theorem), $\exp\left[-\Psi_{(\Lambda,v)}(k)\right]$ is a positive semidefinite function
of $k \in \R$, and equals $1$ at $k=0$.
Therefore, by Bochner's theorem, it is the characteristic function of a unique c.d.f.
We define $F_{(\Lambda,v)}$ to be this c.d.f.
Thus,
$$
\int_{-\infty}^{\infty} e^{ikx}\, dF_{(\Lambda,v)}(x)\, =\, e^{-\Psi_{(\Lambda,v)}(k)}\, .
$$
It is a basic fact that $F_{(\Lambda_1,v_1)} \star F_{(\Lambda_2,v_2)} = F_{(\Lambda_1+\Lambda_2,v_1+v_2)}$.
Therefore, $F_{(\Lambda,v)}$ is infinitely divisible:
in fact $F_{(\Lambda,v)} = F_{(\Lambda/n,v/n)}^{\star n}$.
By the L\'evy-Khinchine formula, specialized to symmetric distributions,
every symmetric, infinitely divisible c.d.f.\
is of this form for a unique pair $(\Lambda,v)$.

Let us suppose that we have a spin glass Hamiltonian, defined as previously
$$
-\sfH_N(\s)\, =\, \sum_{i,j=1}^N \sfJ_N(i,j) \s_i \s_j + h \sum_{i=1}^N \s_i\, .
$$
But, now we suppose that the random couplings, $(\sfJ_N(i,j)\, :\, 1\leq i,j\leq N)$ are i.i.d., and distributed according to $F_N = F(\Lambda/N,v/N)$.
We may denote $F := F_1 = F(\Lambda,v)$.
For a pure Gaussian, with $v=\beta^2/2$, we have $F_N = F(0,\beta^2/2N) = F^*_{N,\beta}$, as before.
For the Poissonized Gaussian, we have $F^*_{N,\alpha,\beta} = F_N = F(\alpha \Lambda^*_{\beta}/N,0)$,
where
$$
\Lambda^*_{\beta}(dy)\, =\, \frac{2 e^{-y^2/\beta^2}}{\sqrt{\pi \beta^2}}\, \boldsymbol{1}_{(0,\infty)}(y)\, dy\, .
$$
Therefore, this does, indeed, generalize the two cases we considered before, of the SK model
and one version of the Viana-Bray model.
Then we write
$$
p^*_N(\Lambda,v)\, =\, \frac{1}{N} \E\left[\ln\left(\sum_{\s \in \Omega_N} e^{-\sfH_N(\s)}\right)\right]\, .
$$
We will not necessarily introduce a new symbol for the Hamiltonian, when the underlying
distribution for the couplings, $(\sfJ_N(i,j)\, :\, 1\leq i,j\leq N)$, changes.
Rather, we will endeavor to write the distribution explicitly, when we take expectations,
as in $\E^{F}[\cdot]$.
Let us write $\E_{(\Lambda,v)}[\cdot]$ instead of $\E^{F_{(\Lambda,v)}}[\cdot]$.

Let us introduce the fundamental definitions of Franz and Leone, Guerra and Toninelli,
and De Sanctis.
First of all, given a function of $n$ spin configurations, $u : (\Omega_N)^n \to \R$,
let us write
$$
\langle u \rangle\, =\, \langle u(\s^{(1)},\dots,\s^{(n)}) \rangle\,
:=\, (\sfZ_n)^{-1} \sum_{\s^{(1)},\dots,\s^{(n)} \in \Omega_N} e^{-[\sfH_N(\s^{(1)})+\dots+\sfH_N(\s^{(n)})]} u(\s^{(1)},\dots,\s^{(n)})\, .
$$
Note that, in our choice of convention, this is still a random variable depending on the underlying
coupling constants.
But $\E_{(\Lambda,v)}[\langle u(\s^{(1)},\dots,\s^{(n)}) \rangle]$ has had the expectation taken
(with respect to the i.i.d.\ product of $F_{(\Lambda,v)}$ distributions).
Let us also define the degree-$n$ multi-overlap function $R_{N,n} : (\Omega_N)^n \to \R$, as
$$
R_{N,n}(\s^{(1)},\dots,\s^{(n)})\, =\, \frac{1}{N} \sum_{i=1}^N \s_i^{(1)} \cdots \s_i^{(n)}\, .
$$
Then a fundamental result of Franz and Leone and Guerra and Toninelli,
suitably generalized to the present context,
gives an integral formula for $p_N^*(\Lambda_2,v_2) - p_N^*(\Lambda_1,v_1)$
in terms of the expectations of these multi-overlaps.
Let us generalize the definition of $a_{2k}(F)$ as follows:
\begin{align*}
a_{0}^*(\Lambda,v)\, &:=\, \frac{v}{2} + \int_{0+}^{\infty} \ln(\cosh(y))\, \Lambda(dy)\, ;\\
a_{2}^*(\Lambda,v)\, &:=\, \frac{v}{2} + \frac{1}{2}\, \int_{0+}^{\infty} \tanh^2(y)\, \Lambda(dy)\, ;\qquad \text{and}\\
a_{2k}^*(\Lambda,v)\,&=\, a_{2k}^*(\Lambda)\,
:= \frac{1}{2k}\, \int_{0+}^{\infty} \tanh^{2k}(y)\, \Lambda(dy)\, ,\qquad \text{for}\qquad k=2,3,\dots\, .
\end{align*}
(In some sense, when we write $(\Lambda,v)$ this really means the distribution $\Lambda + \frac{1}{2} \delta_0''$,
acting on smooth test functions $\phi$ satisfying $\phi(0) = \phi'(0)=0$.)
Then the result is as follows.

\begin{prop}
\label{prop:C}
Suppose that $(\Lambda_1,v_1)$ and $(\Lambda_2,v_2)$ are parameters from the L\'evy-Khinchine formula,
satisfying the further requirement that $a_0(\Lambda_1,v_1)$ and $a_0(\Lambda_2,v_2)$ are finite.
Then
\begin{multline*}
p_N^*(\Lambda_2,v_2) - p_N^*(\Lambda_1,v_1)\,
=\, a_0^*(\Lambda_2,v_2) - a_0^*(\Lambda_1,v_1)\\
- \sum_{k=1}^{\infty} [a_{2k}^*(\Lambda_2,v_2) - a_{2k}^*(\Lambda_1,v_1)]\, \int_0^1
\E_{(t \Lambda_2 + (1-t) \Lambda_1,t v_2 + (1-t) v_1)}\left[\left\langle\left(R_{N,n}(\s^{(1)},\dots,\s^{(n)})\right)^2\right\rangle\right]\, dt\, .
\end{multline*}
\end{prop}

\begin{Remark}
One clearly sees the motivation for Proposition \ref{prop:A} in this formula.
\end{Remark}

\noindent
An immediate corollary, along the lines of Proposition \ref{prop:A} can be deduced from this.
Namely,
$$
|p_N^*(\Lambda_2,v_2) - p_N^*(\Lambda_1,v_1)|\,
\leq\, \sum_{k=0}^{\infty} |a_{2k}^*(\Lambda_2,v_2) - a_{2k}^*(\Lambda_1,v_1)|\, .
$$
This gives a bound which is uniform in $N$, and is often easier to calculate.
Guerra and Toninelli were the first to write this type of bound, when they used it
to give a very simple proof that
$$
\lim_{\alpha \to \infty} \sup_{N \in \Z_{>0}} \left|p^{\rm VB}_N\left(\alpha,\frac{\beta}{\sqrt{2\alpha}}\right)
- p^{\rm SK}_N(\beta)\right|\, =\, 0\, ,
$$
for all $\beta\geq 0$.
This is known as the ``infinite connectivity limit''.

The main advantage of Prop \ref{prop:C} is that it is an exact formula.
For example, Franz and Leone used a close analogue of this formula to
prove that, for the Viana-Bray model, the thermodynamic limit of the pressure
exists.
Guerra and Toninelli used similar methods to control the high-temperature and
low-connectivity regions of phase space, demonstrating replica symmetry in that
domain.
And De Sanctis used that method and other arguments to prove an extended
variational principle, thereby generalizing
the results of Aizenman, Sims, and an author \cite{ASS1},
from the SK to the Viana-Bray model.

A specialized version of Prop \ref{prop:C}, applicable to the standard Viana-Bray model, is contained 
implicitly or explicitly in each of the papers \cite{FranzLeone}, \cite{GT3} and \cite{deSanctis}. Accordingly, the reader may find the
relevant proofs, there. However, a new issue arises in the generalized context. Namely, we should 
prove that the pressure function, $\sfp_N$, is still in the domain of the ``generator'', despite the fact that 
it is not a typical test-function (because it does not vanish at infinity). Next, we will present 
the definition of the generator, as well as this technical result.

\subsection{The generator}
\label{subsec:Levy}
An important fact is that one can define a L\'evy process
associated to the infinitely divisible distribution $F_{(\Lambda,v)}$.
This is a stochastic process $(\sfX_t\, :\, t \geq 0)$.
Among other properties are these two:
for each $s,t \geq 0$, the increment $(\sfX_{s+t}-\sfX_s)$ is independent
of $\calF_s$, where $\calF_s$ is the $\s$-algebra generated by $(\sfX_{r}\, :\, 0\leq r\leq s)$;
and $(\sfX_{s+t}-\sfX_s)$ has the c.d.f.\ $F_{(t\Lambda,tv)}$.
In particular, the increments are independent and stationary.
(There are also continuity properties of the L\'evy process which we will not need.)
See \cite{Sato}, for example, for a reference.
Let $C^2_0(\R)$ denote the set of function $f \in \R$ which are twice continuously
differentiable, and such that
$$
\lim_{|x| \to \infty} f(x)\,
=\, \lim_{|x| \to \infty} f'(x)\,
=\, \lim_{|x| \to \infty} f''(x)\,
=\, 0\, .
$$
Then the following is a specialization of Theorem 31.5 in \cite{Sato}:
If $f \in C^2_0(\R)$, then $\E[f(\sfX_t)]$ is differentiable, and
\begin{equation}
\label{eq:genderiv}
\frac{d}{dt} \E[f(\sfX_t)]\, =\, \E[\mathfrak{G}_{(\Lambda,v)} f(\sfX_t)]\, ,
\end{equation}
where $\mathfrak{G}_{(\Lambda,v)}$ is the generator
$$
\mathfrak{G}_{(\Lambda,v)} f(x)\,
=\, \frac{v}{2} f''(x) +  \int_{0+}^{\infty}
\left(\frac{1}{2} f(x+y) - f(x) + \frac{1}{2} f(x-y)\right)\, \Lambda(dy)\, .
$$
In particular, if we instead consider $\sfX$ to be distributed by $F_{(\Lambda,v)}$, and denote
the expectation with respect to $F_{(\Lambda,v)}$ as $\E_{(\Lambda,v)}$,
then we have
$$
\frac{d}{dt} \E_{(t \Lambda, t v)}[f(\sfX)]\, =\, \E_{(t \Lambda, t v)}[\mathfrak{G}_{(\Lambda,v)} f(\sfX)]\, .
$$

The technical fact we want to prove now is the following:

\begin{lemma}
\label{lem:generalGenerator}
Suppose $(\Lambda_1,v_1)$ and $(\Lambda_2,v_t)$ satisfy $a_0^*(\Lambda_1,v_1)<\infty$
and $a_0^*(\Lambda_2,v_2) < \infty$.
Also suppose that $f : \R \to \R$ is a function in $C^2(\R)$ such that $f'$ and $f''$ are in $L^{\infty}(\R)$.
(In particular, then, $f$ is globally Lipschitz.)
Then $\mathfrak{G}_{(\Lambda_1,v_1)} f$ and $\mathfrak{G}_{(\Lambda_2,v_2)} f$
are both well-defined and
\begin{equation}
\label{eq:tech}
\E_{(\Lambda_2,v_2)}[f(\sfX)] - \E_{(\Lambda_1,v_1)}[f(\sfX)]\,
=\, \int_0^1 \E_{(t \Lambda_2 + (1-t) \Lambda_1,t v_2 + (1-t) v_1)}\left[\left(\mathfrak{G}_{(\Lambda_2,v_2)}-\mathfrak{G}_{(\Lambda_1,v_1)}\right) f(\sfX)\right]\, dt\, .
\end{equation}
\end{lemma}

This lemma proves that the generator is applicable even to the function $\sfp_N$,
which is a Lipschitz function of the random coupling constants $\sfJ(i,j)$.
The beautiful calculation of the generator for this function can be read off of any
of the references mentioned before, \cite{FranzLeone,GT3,deSanctis}.
We will spend the rest of this section proving Lemma \ref{lem:generalGenerator}.

\subsection{Proof}
First, supposing $a_0^*(\Lambda,v)<\infty$, let us show that $F^*_{(\Lambda,v)}$
has finite first moment.
This is equivalent to checking that $\ln(\cosh(x))$ is integrable, because $\ln(\cosh(x)) \sim |x|$
as $|x| \to \infty$.
But $\ln(\cosh(x)) - \epsilon^{-1} \ln(\cosh(\epsilon x))$ is a nonnegative function, for every $0<\epsilon<1$,
and it is in $C^2_0(\R)$.
So the generator applies to it.
But it is easy to see that, defining $u_{\epsilon}(x) = \epsilon^{-1} \ln(\cosh(\epsilon x))$,
$$
\mathfrak{G}_{(\Lambda,v)} u_{\epsilon}(x)\, =\, \frac{\epsilon v}{2 \cosh^2(\epsilon x)}
+ \frac{1}{2\epsilon} \int_{0+}^{\infty} \ln\left(1 + \frac{\sinh^2(\epsilon y)}{\cosh^2(\epsilon x)}\right)\, \Lambda(dy)\, .
$$
There are several things to note. First, the formula is well-defined.
Second, an upper bound is obtained by bounding $\cosh^2(\epsilon x)\geq 1$
and $(2\epsilon)^{-1} \ln(1+\sinh^2(\epsilon y)) = \epsilon^{-1} \ln(\cosh(\epsilon y)) \leq \ln(\cosh(y))$.
Therefore, $\mathfrak{G}_{(\Lambda,v)} u_{\epsilon}(x) \leq a_0^*(\Lambda,v)$, for all $x$.
But finally,  by the DCT, $\lim_{\epsilon \to 0} \mathfrak{G}_{(\Lambda,v)} u_{\epsilon}(x) \to 0$,
for all $x$.
So, by various trivial applications of the DCT,
\begin{align*}
\E_{(\Lambda,v)}\left[\ln(\cosh(\sfX))\right]\,
=\ \lim_{\epsilon \to 0} \E_{(\Lambda,v)}\left[(u_1-u_{\epsilon})(\sfX)\right]\,
&=\, \lim_{\epsilon \to 0} \int_0^1 \E_{(t\Lambda,t v)}\left[\mathfrak{G}_{(\Lambda,v)}(u_1-u_{\epsilon})(\sfX_t)\right]\, dt\\
&=\, \int_0^1 \E_{(t\Lambda,t v)}\left[\mathfrak{G}_{(\Lambda,v)}u_1(\sfX_t)\right]\, dt\\
&\leq\, a_0^*(\Lambda,v)\, ,
\end{align*}
A quantitative version shows that $\E_{(\Lambda,v)}[|\sfX|] \leq (1+ a_0^*(\Lambda,v)) e$.
In particular, by assumption $a_0^*(\Lambda_1,v_1)$ and $a_0^*(\Lambda_2,v_2)$ are finite,
therefore, $F_{(t\Lambda_1+(1-t)\Lambda_2,t v_1 + (1-t) v_2)}$ has a finite first moment for all $t \in [0,1]$,
and moreover it is bounded by $(1+\max\{a_0^*(\Lambda_1,v_1),a_0^*(\Lambda_2,v_2)\})e$.
Since $f$ is globally Lipschitz, this means it is integrable against $F_{(t\Lambda_1+(1-t)\Lambda_2,t v_1 + (1-t) v_2)}$.

Now we simply introduce a smooth cut-off, and perform basic estimates.
Suppose that $\psi(x)$ is any function which is twice continuously differentiable,
compactly supported, and such that $\psi(0)=1$.
Then, defining, $f_{\epsilon}(x) = f(x) \psi(\epsilon x)$ this is in $C^2$ with compact support.
So equation (\ref{eq:tech}) holds with $f$ replaced by $f_{\epsilon}$.
Since $f_{\epsilon} \to f$, pointwise, and since $|f_{\epsilon}(x)|\leq \|\psi\|_{\infty} |f(x)|$, we have
$$
\E_{(\Lambda_2,v_2)}[f(\sfX)] - \E_{(\Lambda_1,v_1)}[f(\sfX)]\,
=\, \lim_{\epsilon \to 0} \left(\E_{(\Lambda_2,v_2)}[f_{\epsilon}(\sfX)] - \E_{(\Lambda_1,v_1)}[f_{\epsilon}(\sfX)]\right)\, .
$$
To prove (\ref{eq:tech}) for $f$ (instead of $f_{\epsilon}$), we just have to show that the right hand side of this equation converges to the right hand
side of (\ref{eq:tech}) as $\epsilon \to 0$.
Note that
$$
|f_{\epsilon}(x) - f_{\epsilon}(y)|\, \leq\, (\|f'\|_{\infty} + \epsilon \|\psi'\|_{\infty}) |x-y|\, .
$$
So, for some $A<\infty$,
$$
\left|\frac{1}{2} f_{\epsilon}(x+y) + \frac{1}{2} f_{\epsilon}(x-y) - f_{\epsilon}(x)\right|\, \leq\, A |y|\, ,
$$
which we use for $|y|\geq 1$.
Also,
$$
\|f_{\epsilon}''(x)\|
\leq\,\|f''\|_{\infty} \|\psi\|_{\infty} + 2 \epsilon \|f'\|_{\infty} \|\psi'\|_{\infty} + \epsilon^2 \|\psi''\|_{\infty} |f(x)|\, .
$$
So, by Taylor's theorem, for some $B,C<\infty$,
$$
\left|\frac{1}{2} f_{\epsilon}(x+y) + \frac{1}{2} f_{\epsilon}(x-y) - f_{\epsilon}(x)\right|\, \leq\, (B + \epsilon C |f(x)|) y^2\, ,
$$
which we use for $|y|<1$.
So, since $k \ln(\cosh(y)) \leq \min(y^2,|y|) \leq K \ln(\cosh(y))$, for some $0<k<K<\infty$, we have
$$
\left|\frac{1}{2} f_{\epsilon}(x+y) + \frac{1}{2} f_{\epsilon}(x-y) - f_{\epsilon}(x)\right|\, \leq\, (\tilde{B} + \epsilon \tilde{C} |f(x)|) \ln(\cosh(y))\, ,
$$
for some different constants $\tilde{B}, \tilde{C} < \infty$,
and all $y\in\R$.
Since $a_0^*(\Lambda_1,v_1)$ and $a_0^*(\Lambda_2,v_2)$ are both finite, this upper bound is integrable against $\Lambda_1$
and $\Lambda_2$.
Therefore, for some other constants $\hat{A}, \hat{B}<\infty$,
\begin{equation}
\label{ineq:Gineq}
|\mathfrak{G}_{(\Lambda_i,v_i)} f_{\epsilon}(x)|\, \leq\, \hat{A} + \hat{B} |f(x)|\qquad \text{for}\qquad i=1,2\, .
\end{equation}
But also note that,
$$
\lim_{\epsilon \to 0} \frac{1}{2} f_{\epsilon}(x+y) + \frac{1}{2} f_{\epsilon}(x-y) - f_{\epsilon}(x)\, =\,
\frac{1}{2} f(x+y) + \frac{1}{2} f(x-y) - f(x)\, ,
$$
for each $x\in \R$ and $y>0$. Similarly, $\lim_{\epsilon \to 0} f_{\epsilon}''(x) = f''(x)$, for each $x \in \R$.
So, by the DCT, we know
$$
\lim_{\epsilon \to 0} \mathfrak{G}_{(\Lambda_i,v_i)} f_{\epsilon}(x)\,
=\, \mathfrak{G}_{(\Lambda_i,v_i)} f(x)\, ,
$$
for $i=1,2$ and all $x \in \R$.
But also, the upper bound of (\ref{ineq:Gineq}) is integrable against $F_{(t\Lambda_1+(1-t)\Lambda_2,t v_1 + (1-t) v_2)}$,
for all $t$, and is independent of $\epsilon$.
So, by the DCT again,
\begin{multline*}
\lim_{\epsilon \to 0}
\E_{(t \Lambda_2 + (1-t) \Lambda_1,t v_2 + (1-t) v_1)}\left[\left(\mathfrak{G}_{(\Lambda_2,v_2)}-\mathfrak{G}_{(\Lambda_1,v_1)}\right) f_{\epsilon}(\sfX)\right]\\
=\, \E_{(t \Lambda_2 + (1-t) \Lambda_1,t v_2 + (1-t) v_1)}\left[\left(\mathfrak{G}_{(\Lambda_2,v_2)}-\mathfrak{G}_{(\Lambda_1,v_1)}\right) f(\sfX)\right]\, .
\end{multline*}
Of course, the integrated version of this is also true, once again by DCT.

\section{Conclusion}
We considered two different types of bounds for the difference of
two pressures of two spin glasses. In Section 1, we considered a
bound which demonstrates that the pressure is Lipschitz with respect
to a seminorm, $\|F\| = \sum_{k=0}^{\infty} |a_{2k}(F)|$. This was
strongly motivated by Carmona and Hu's proof of universality for the
SK model, but generalized to also apply to the Viana-Bray model. In
Section 2, we considered a bound which is useful if one does not
assume that the coupling constants are independent. This proved a
different type of universality, which was noted, but not proved, in
papers of Franz and Leone, and Guerra and Toninelli. In Section 3,
we briefly reviewed the theory of infinitely divisible
distributions, and applied it to the Viana-Bray model. All the
results in this letter are simple. But we hope they add something to
the growing wealth of knowledge for mean-field spin glass models.

\section*{Acknowledgements}

The research of 
S.S.\ was supported in part by a U.S.\ National Science Foundation
grant, DMS-0706927.

\end{document}